\documentclass[osajnl,twocolumn,showpacs,superscriptaddress,10pt]{revtex4-1} 
\usepackage{amsmath,amssymb,graphicx}

\usepackage{mathtools}
\usepackage{xfrac}

\begin{document}

\title{Coherent two-octave-spanning supercontinuum generation in lithium-niobate waveguides}

\author{Mengjie Yu}\email{Corresponding author: mjyu@g.harvard.edu}
\affiliation{John A. Paulson School of Engineering and Applied Sciences, Harvard University, Cambridge, MA 02138}

\author{Boris Desiatov}
\affiliation{John A. Paulson School of Engineering and Applied Sciences, Harvard University, Cambridge, MA 02138}

\author{Yoshitomo Okawachi}
\affiliation{Department of Applied Physics and Applied Mathematics, Columbia University, New York, NY 10027}

\author{Alexander L. Gaeta}
\affiliation{Department of Applied Physics and Applied Mathematics, Columbia University, New York, NY 10027}

\author{Marko~Lon\v{c}ar}
\affiliation{John A. Paulson School of Engineering and Applied Sciences, Harvard University, Cambridge, MA 02138}

\begin{abstract}We demonstrate coherent supercontinuum generation (SCG) in a monolithically integrated lithium-niobate waveguide, under the presence of second- and third-order nonlinear effects. We achieve more than two octaves of optical bandwidth in a 0.5-cm-long waveguide with 100-picojoule-level pulses. Dispersion engineering of the waveguide allows for spectral overlap between the SCG and the second harmonic which enables direct detection of the carrier-envelope offset frequency $f\textsubscript{CEO}$ using a single waveguide. We measure the $f\textsubscript{CEO}$ of our femtosecond pump source with a 30-dB signal-to-noise ratio.    
\end{abstract}


\maketitle 


The generation of a coherent supercontinuum spectrum with modelocked laser pulses is critical for many frequency-comb-based applications, including time and frequency metrology, optical frequency synthesis, microwave generation, all-optical clocks, and spectroscopy \cite{Diddams}. A coherent octave-spanning supercontinuum spectrum allows for the detection of the carrier-envelope offset frequency $f\textsubscript{CEO}$ through self-referenced $f$-$2f$ interferometry \cite{Telle, Diddams00, Dudley}, allowing for a stabilized frequency comb. Over the past decade, there has been significant advances in nanofabrication technology that have led to the development of various chip-based platforms for supercontinuum generation (SCG) \cite{Kuyken,Duchesne, Halir, Epping,Leo,Kuyken15,Singh15,Mayer,Johnson,Klenner,Liu,Oh,Porcel,Carlson,Hickstein,Okawachi,Singh,Guo,Sinobad,Okawachi18,Hickstein18}. Previously, $f\textsubscript{CEO}$ detection and stabilization was demonstrated in silicon nitride \cite{Mayer,Klenner,Carlson,Hickstein,Okawachi18,Hickstein18}, and more recently, simultaneous SCG and second harmonic generation (SHG) has allowed for on-chip $f$-$2f$ interferometry \cite{Okawachi18,Hickstein18}. However since silicon nitride is a centrosymmetric material, it is only possible to achieve modest second-harmonic efficiencies. Alternatively, simultaneous SCG and SHG has been demonstrated in aluminum nitride using 800-pJ pulse energies \cite{Hickstein}. Lithium niobate (LN, LiNbO$_3$) is a promising nonlinear material for realizing an $f$-$2f$ interferometer as it possesses a large nonlinear index ($n_2$ = 2.5 $\times$ 10$^{-19}$ m$^2$/W) and has a stronger $\chi^{(2)}$ nonlinearity ($r_{33}$ = 3 $\times$ 10$^{-11}$ m/V) \cite{Nikogosyan,Guarino,Wolf,Rao,Kowligy,He}. Previous demonstrations have shown supercontinuum spectra exceeding an octave in periodically-poled lithium niobate (PPLN) \cite{Phillips,Iwakuni,Kowligy}. However, these devices typically suffer from low index contrast between the core and the cladding resulting in low optical confinement and large device dimensions and requires > 1 nJ pulse energies. Recently, Zhang, $et$ $al.$ \cite{ZhangOptica}, has demonstrated ultralow-loss monolithically integrated LN waveguides with 2.7 dB/m propagation loss using thin-film LN-on-insulator technology. This platform has been leveraged to realize integrated PPLN devices, Kerr-combs and electro-optic combs \cite{WangOptica,WangArXiv,Zhang}. 

   \begin{figure}[h!]
\centerline{\includegraphics[width=9cm]{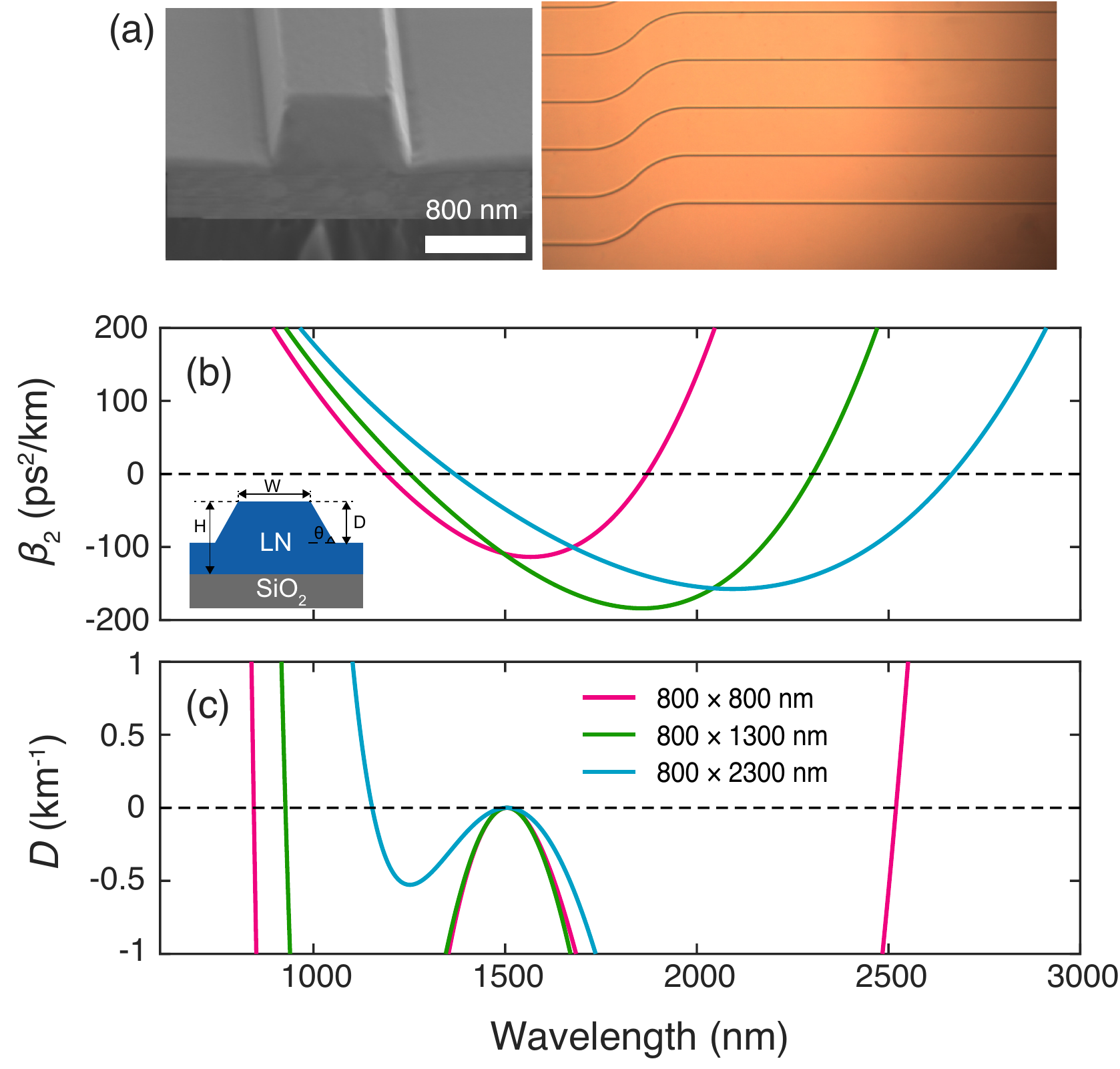}}
\caption{(a) Left: Scanning electron microscope (SEM) picture of the air-clad LN waveguide cross section (top width: 800 nm). Right: Microscope image of the waveguide. (b) Simulated group-velocity dispersion (c) the corresponding dispersion operator for a 1506 nm pump wavelength for three different cross sections (H $\times$ W) , 800$\times$800 nm, 800$\times$1300 nm, and 800$\times$2300 nm. The spectral position of the dispersive wave can be predicted by the zero-crossings of the dispersion operator $\it{D}$. Embedded: H, film thickness: 800 nm; W, top width; D, etch depth: 450 nm; $\theta$, sidewall angle: 60 degree.}
\label{Fig1}
\end{figure}

In this paper, we build on these results and demonstrate coherent SCG in a monolithically integrated LN waveguide that spans over two octaves using picojoule-level pulses. Our work represents the broadest supercontinuum spectrum generated, to our knowledge, in an integrated LN waveguide with a total bandwidth spanning 2.58 octaves from 400 to 2400 nm. In addition, the high $\chi^{(2)}$ effect in LN waveguides allows for efficient SHG. We theoretically and experimentally investigate the dispersion in LN waveguides and perform dispersion engineering to allow for spectral overlap between the SCG and second harmonic signal, allowing for $f_{\text{CEO}}$ detection in the same device with a 30-dB signal-to-noise ratio of the beatnote. Our results illustrate the potential of LN as an efficient nonlinear photonic platform for chip-scale SCG. 

The LN devices are fabricated using 800-nm-thick X-cut LN thin films on top of a 2-$\mu$m silicon dioxide (SiO$_2$) layer on silicon substrates (NanoLN). First, the photonic structures are patterned with electron beam lithography (EBL) using the Elionix ELS-F125 tool with a HSQ resist. Next, the patterns are transferred onto the LN thin film using an optimized Ar+ plasma etching recipe in the reactive ion etching (RIE) tool. Finally, the residual EBL mask is removed by a buffered oxide etch (BOE). Figure \ref{Fig1}(a) shows a scanning electron micrograph (SEM) image of an air-cladded LN waveguide. The propagation loss of the LN waveguides is estimated to be 3 dB/m.

To allow for broadband coherent SCG, we examine dispersion engineering of the LN waveguide \cite{Turner}. We simulate the dispersion of the waveguide using a finite-element mode solver. Figure \ref{Fig1}(b) shows the simulated group-velocity dispersion (GVD) for the fundamental transverse electric (TE) mode of three different waveguide top widths of 800, 1300 and 2300 nm, while Fig. \ref{Fig1}(c) shows the corresponding dispersion operator $\it{D}$ for a pump centered at 1506 nm. The operator is expressed as \cite{Dudley,Okawachi},

\begin{equation}
D = \sum_{\mathclap{n=2,3,...}}\frac{\beta_n(\omega_0)}{n!}(\omega-\omega_0)^n,
\end{equation}

\noindent where $\beta_n$ corresponds to the $\it{n}$-th order dispersion coefficient, and $\omega_0$ is the pump frequency. The spectral position where the dispersive wave (DW) occurs can be predicted from the around-zero-crossing of the dispersion operator. To allow for spectral overlap between the supercontinuum and the second harmonic signal, we choose the cross section 800$\times$800 nm. 

\begin{figure}[t!]
\centerline{\includegraphics[width=9cm]{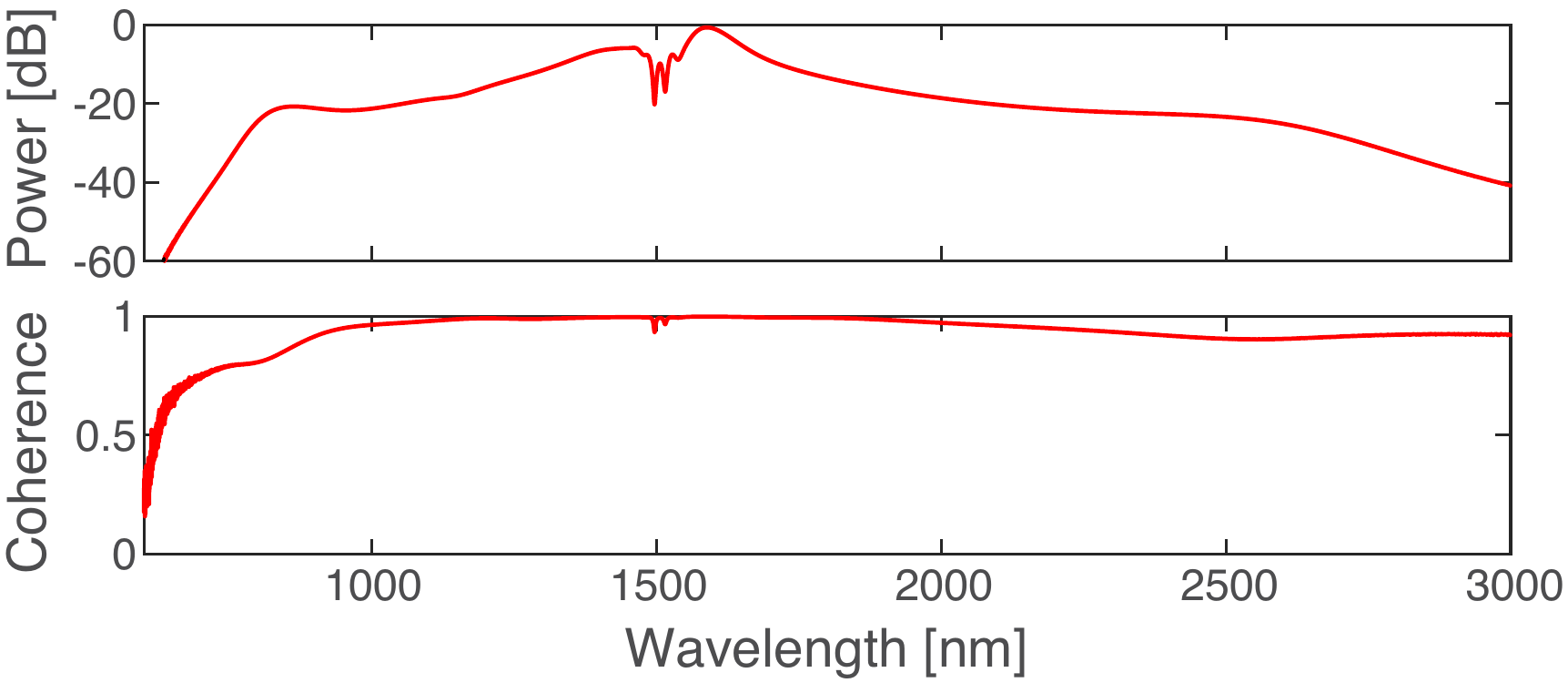}}
\caption{Calculated average SCG spectrum (top) and the first-order mutual coherence (bottom) based on 128 individual simulations. The LN waveguide is 0.5 cm long with a cross section of 800$\times$800 nm.}
\label{Fig2}
\end{figure}

Next, we theoretically consider the propagation dynamics in LN waveguides. We simulate the generated spectrum by numerically solving the nonlinear envelope equation using the split-step Fourier method \cite{Gaeta} with the inclusion of third-order nonlinearity, higher-order dispersion, and self-steepening. We neglect the contributions from $\chi^{(2)}$ and Raman to isolate DW generation. The LN waveguide is 0.5 cm long with a cross section of 800$\times$800 nm. We use 160-fs pulses with a pulse energy of 187 pJ in the waveguide (peak power = 1.17 kW). To characterize the coherence, we calculate the first-order mutual coherence function \cite{Gu,Ruehl} by simulating 128 individual spectra where the input pulses are seeded with simulated quantum noise. Figure \ref{Fig2}(top) shows the averaged supercontinuum spectrum which shows a DW near 800 nm in agreement with the dispersion operator prediction in Fig. \ref{Fig1}. Figure \ref{Fig2}(bottom) shows the calculated coherence function, and we verify that the coherence is near unity over most of the generated spectrum.     

\begin{figure}[h!]
\centering
\centerline{\includegraphics[width=9cm]{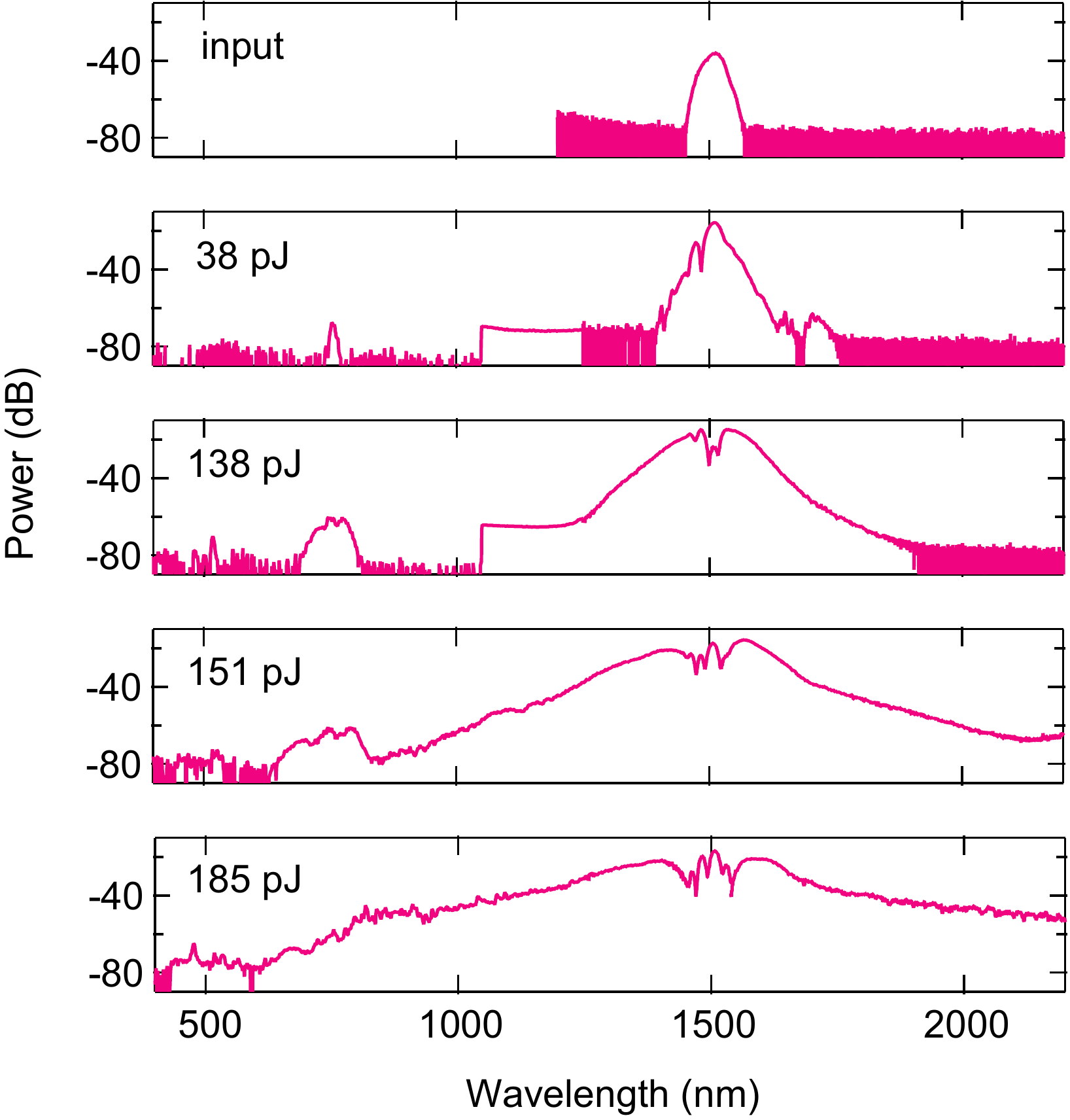}}
\caption{Measured spectra for the input and various pump pulse energies in the waveguide of 38 pJ, 138 pJ, 151 pJ, and 185 pJ (top to bottom). The input coupling loss is 8.5 dB. We have not observed material damage up to 7-kW peak power. The spectrum shows various nonlinear interactions due to both $\chi^{(2)}$ and $\chi^{(3)}$ nonlinearities, including supercontinuum generation, second harmonic generation, sum frequency generation and third-order harmonic generation.}
\label{Fig3}
\end{figure}

In our experiment, we send pulses from a femtosecond optical parametric oscillator centered at 1506 nm with an 80-MHz repetition rate into a monolithically integrated air-clad LN waveguide. The LN waveguide is fabricated to be 0.5 cm long with a top width of 800 nm, similar to the conditions of the previous simulation. The pump is launched along the y-axis of LN thin-film crystal at the fundamental TE mode of the waveguide, which has the strongest $\chi^{(2)}$ component. The output is collected using a lensed fiber and sent to three different optical spectrum analyzers covering 400 -- 2500 nm. We measure the coupling loss to be 8.5 dB at the input facet. Figure \ref{Fig3} shows the input spectrum along with the generated SCG spectra for various pulse energies in the waveguide which are extrapolated from the input coupling loss. At 38 pJ of pulse energy, we observe the onset of SHG at 750 nm and two spectral features near 1660 nm and 1730 nm which we attribute to Raman scattering corresponding to the optical phonon modes of A(TO$_4$) and A(LO$_4$), respectively. As the pump pulse energy is further increased, we observe further spectral broadening around the pump, which is largely due to self-phase modulation, and broadening of the SHG component. At 151 pJ, SCG and the SHG components begin to spectrally overlap, and we observe broadening of a spectral component around 500 nm due to sum frequency generation (SFG) between the SCG and SHG components. Finally, at 185 pJ of pulse energy, both SHG and SFG cover the entire visible spectrum, and together with SCG, span 2.58-octaves of bandwidth (630THz) from 400 nm to 2400 nm [Fig.4 (top)]. Moreover, the DW is generated around 800 nm which merges with the broadened SHG component.

\begin{figure}[h!]
\centerline{\includegraphics[width=9cm]{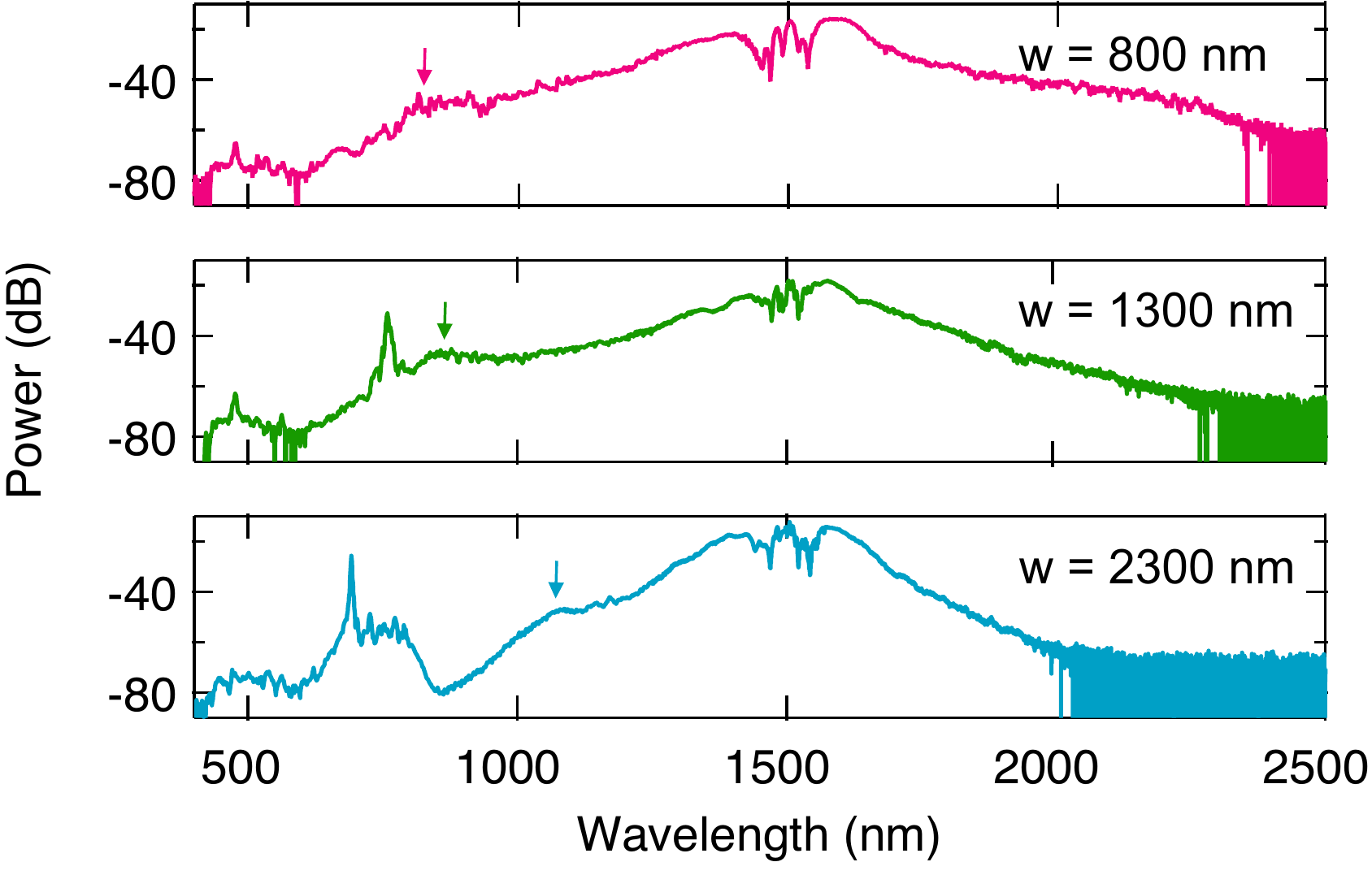}}
\caption{Experimental spectra for three different cross sections 800$\times$800 nm, 800$\times$1300 nm, and 800$\times$2300 nm. The arrow indicates the generated DW shifts towards the pump wavelength with increasing waveguide width, which agrees with our prediction in Fig. \ref{Fig1}.}
\label{Fig4}
\end{figure}

We experimentally verify the ability to dispersion engineer in integrated LN waveguides by investigating three different waveguide cross sections, 800$\times$800 nm, 800$\times$1300 nm, and 800$\times$2300 nm, and Fig. \ref{Fig4} shows the generated SCG spectra. We observe that the DW spectrally shifts to longer wavelengths as the waveguide width is increased, which agrees well with the dispersion operator predictions in Fig \ref{Fig1} (c). The SCG of the 800-nm-width waveguide yields better overlap with SHG as compared to the other two widths, and its spectral bandwidth extends further due to the second DW as shown in Fig \ref{Fig1} (c). The discrepancy in the spectrum at high wavelengths compared to our simulation can be attributed to linear loss and wavelength-dependent coupling loss of the collection silica lensed fiber. In addition, for the 800$\times$1300 nm, and 800$\times$2300 nm widths, we observe a sharp SHG peak which we attribute to phase matching with a 4th-order spatial mode at visible wavelengths \cite{Okawachi18}. The difference in the position of the DW compared to the dispersion operator prediction in Fig. \ref{Fig1} can be attributed to the deviation of the actual waveguide dimension from the simulation due to waveguide fabrication tolerances. 

\begin{figure}[t]
\centerline{\includegraphics[width=6cm]{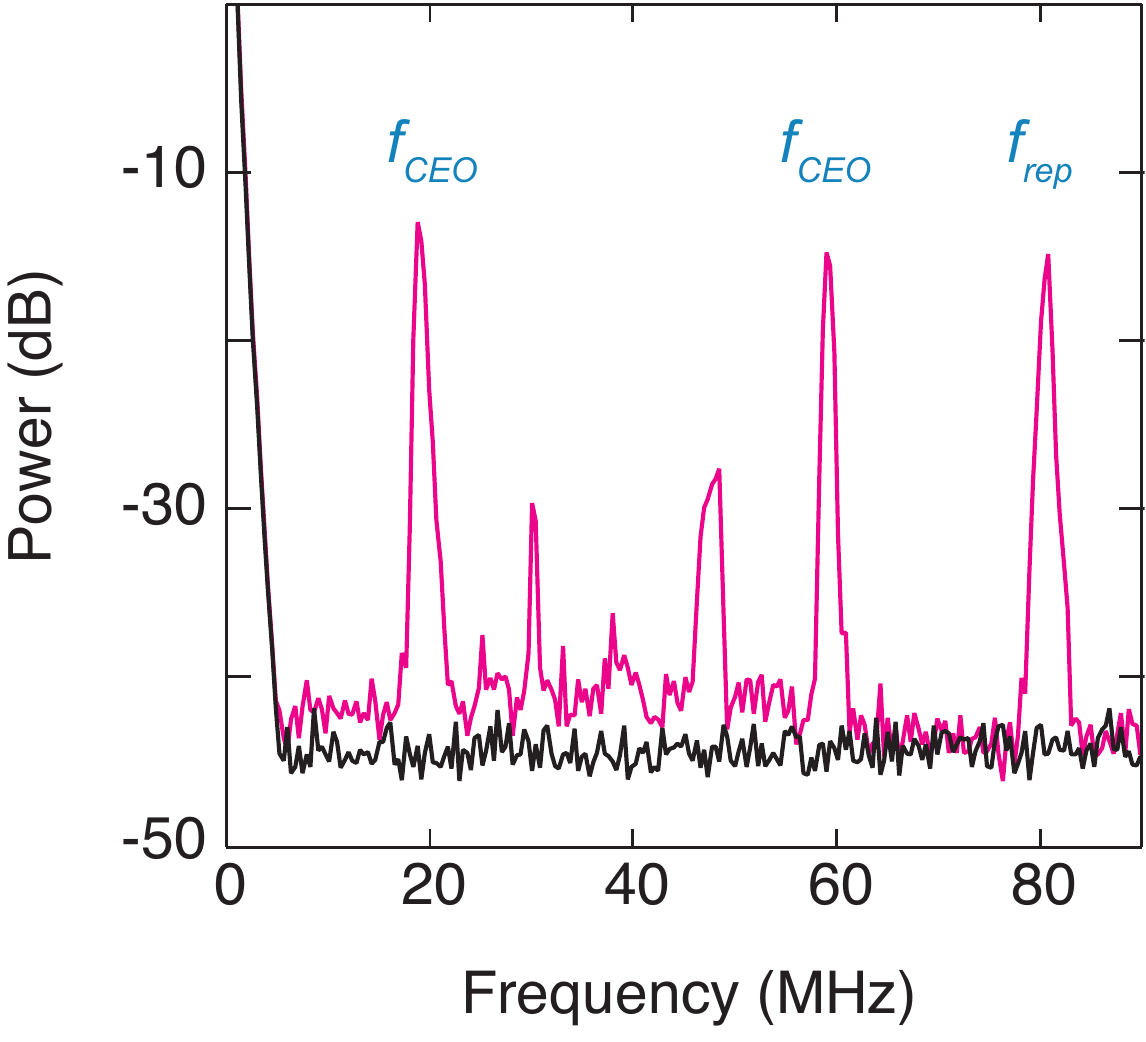}}
\caption{Measured RF spectrum (magenta). We measure a $f_{\text{CEO}}$ beatnote with a signal-to-noise ratio of 30 dB at 1-MHz resolution bandwidth. The black curve indicates the detector background.  } 
\label{Fig5}
\end{figure}

Finally, we directly measure the $f_{\text{CEO}}$ beatnote by measuring the spectral component near 800 nm where the SCG and SHG components spectrally overlap. For our measurement, we filter the generated supercontinuum using a shortpass filter with a cut-off wavelength of 900 nm and detect the RF spectrum using an avalanche photodiode (Thorlabs APD120A) and an RF spectrum analyzer. Figure \ref{Fig5} shows that the $f_{\text{ref}}$ is at 80 MHz, and the $f_{\text{CEO}}$ and $f_{\text{ref}}$ - $f_{\text{CEO}}$ is 20 and 60 MHz (both labelled $f_{\text{CEO}}$ due to indistinguishability). The other two peaks at 30 and 50 MHz are still under investigation. We measure a $f_{\text{CEO}}$ beatnote with a 30-dB signal-to-noise ratio (SNR) at a 1-MHz resolution bandwidth. The high SNR is enabled due to the spectral overlap between the DW and the SHG signal. Efforts are ongoing to further improve the coupling and propagation losses to allow for efficient SCG and enhancement of the SNR of the $f_{\text{CEO}}$ signal.

In conclusion, we demonstrate a 2.58-octave-spanning supercontinuum in a monolithically integrated LN waveguide. Dispersion engineering of the LN waveguide allows for high spectral overlap between the SCG and SHG components allowing for $f_{\text{CEO}}$ beat detection with a high SNR of 30 dB. By increasing the waveguide length to 2-cm, the required pump energy could be reduced to 30 pJ while maintaining the spectral coherence. Alternatively, by using a monolithically integrated PPLN waveguide, we can further enhance the SHG efficiency and the third order nonlinearity through the cascaded $\chi^{(2)}$ effect. Since the electro-optic effect is dispersive, electrical tuning can be explored for the second harmonic conversion efficiency. Our results offer promise for the development of a monolithic LN photonic platform for chip-scale optical clocks. 

\textbf{Funding.} National Science Foundation (NSF) (ECCS1609549, ECCS-1740296 E2CDA); Defense Advanced Research Projects Agency (DARPA) (W31P4Q-15-1-0013); Air Force Office of Scientific Research (AFOSR) (FA9550-15-1-0303).
\\
\\
\textbf{Acknowledgment.} Device fabrication is performed at the Harvard University Center for Nanoscale Systems (CNS), a member of the National Nanotechnology Coordinated Infrastructure Network (NNCI), which is supported by the National Science Foundation under NSF ECCS award no.1541959. M.Y. and B.D. contributed equally to this work.


\clearpage
\newpage

\end{document}